\documentclass[aps,prl,preprint,superscriptaddress]{revtex4}

\usepackage{graphicx}  
\usepackage{dcolumn}   
\usepackage{bm}        
\usepackage{amssymb}   
\usepackage{color}
\usepackage{amsmath}
\usepackage{epstopdf}
\usepackage{natbib} 
\usepackage{float}

\graphicspath{{Figures/}} 

\def\degrees{$^{\circ}$}

\begin{document}
\title{High-frequency acoustic droplet vaporization\\ is initiated by resonance}
\author{Guillaume Lajoinie\footnotemark}
\affiliation{Physics of Fluids Group, MESA+ Institute for Nanotechnology, Technical Medical (TechMed) Center, University of Twente, P.O. Box 217, 7500 AE Enschede, The Netherlands}       
\author{Tim Segers\footnotemark[\value{footnote}] \footnotetext{Both authors contributed equally to this manuscript}}
\affiliation{Physics of Fluids Group, MESA+ Institute for Nanotechnology, Technical Medical (TechMed) Center, University of Twente, P.O. Box 217, 7500 AE Enschede, The Netherlands}       
\author{Michel Versluis}
\affiliation{Physics of Fluids Group, MESA+ Institute for Nanotechnology, Technical Medical (TechMed) Center, University of Twente, P.O. Box 217, 7500 AE Enschede, The Netherlands}       

\begin{abstract}
Vaporization of low-boiling point droplets has numerous applications in combustion, process engineering and in recent years, in clinical medicine. However, the physical mechanisms governing the phase conversion are only partly explained. Here, we show that an acoustic resonance can arise from the large speed of sound mismatch between a perfluorocarbon microdroplet and its surroundings. The fundamental resonance mode obeys a unique relationship $kR$~$\sim$ 0.65 between droplet size and driving frequency that leads to a 3-fold pressure amplification inside the droplet. Classical nucleation theory shows that this pressure amplification increases the nucleation rate by several orders of magnitude. These findings are confirmed by high-speed imaging performed at a timescale of ten nanoseconds. The optical recordings demonstrate that droplets exposed to intense acoustic waves generated by inter-digital-transducers nucleate only if they match the theoretical resonance size.
\end{abstract}

\maketitle

Vaporization of low-boiling point droplets is omnipresent in today's society with applications in renewable energy and energy storage~\cite{Delgado2012}, combustion~\cite{Zhang2019a}, intumescent fire-protective coatings~\cite{Alongi2015}, and recently in clinical medicine~\cite{Wilson2013,Sheeran2012a}. Deterministic vaporization can be initiated by heat or negative pressure, {and by combinations thereof~\cite{Vinogradov2008}}, which allows droplet vaporization to be triggered by laser light~\cite{Dove2014,Lajoinie2020}, ultrasound~\cite{Kripfgans2000}, {neutrons~\cite{Greenspan1967},} and protons~\cite{Carlier2020}. Ultrasound-triggered phase-change of superheated nano- and microdroplets is known as acoustic droplet vaporization (ADV)~\cite{Williams2013, Couture2006,Sheeran2012b, Reznik2012}. ADV is of great interest for medicine since submicrometer-sized surfactant-stabilized droplets, or nanodroplets, have been shown to be able to extravasate leaky tumor vasculature thereby passively accumulating within the tumor~\cite{Rapoport2007, Mohan2010}. Upon phase change, the formed bubbles can perform therapeutic action such as local drug delivery and sonoporation~\cite{Fix2017,Adan2012,Fabiilli2010}. 

The interaction of ultrasound with a low-boiling point droplet has been subject to extensive study aiming at understanding the underlying physical mechanisms driving nucleation. 
Experimentally, it has been found that a prominent {peak negative pressure (PNP)} nucleation threshold exists above which the nucleation probability increases with the acoustic pressure amplitude and that this threshold, counterintuitively, lowers with an increase in ultrasound frequency and with a decrease in ambient pressure~\cite{Kripfgans2000, Williams2013,Giesecke2003,Schad2010,Aliabouzar2019,Kao2014,Rojas2019,Lin2017,Lu2019,Chattaraj2016,Capece2016,Zhang2019}. These observations are in line with what is predicted from both classical nucleation theory~\cite{Raut2019} and superhamonic focusing~\cite{Shpak2014,Aliabouzar2019}, and by the combination of the two~\cite{Miles2016, Miles2018}. Superharmonic focusing results from the focusing of higher harmonics with wavelengths on the order of the droplet diameter that are generated through nonlinear propagation of the transmitted ultrasound wave~\cite{Shpak2014}. However, the required nonlinear propagation in tissue is dramatically lower from that in water due to the two orders of magnitude lower ratio of acoustic nonlinearity to attenuation, or Gol'dberg number~\cite{goldberg1957}. This severely limits the effectivity of ADV by superharmonic focusing \emph{in vivo}. In this Letter, a physical ADV nucleation mechanism based on an acoustic resonance of the droplet is presented that has been overlooked until now. The theoretical resonance behavior is experimentally validated and its role in lowering the vaporization threshold is elucidated using classical nucleation theory. Resonant ADV offers an efficient approach for in vivo applications and adds to our fundamental understanding of acoustic droplet vaporization.

Here we will give the main results of the derivation; all details can be found in the Supplementary Information SI.1. The system consists of a perfluoropentane (PFP) droplet (medium~1) immersed in water (medium 0) considered to be of infinite size. We assume a purely spherical geometry. To calculate the resonance behavior we couple the pressure on the inside of the droplet interface $p(R_{in},t)$ to the external acoustic driving pressure $p_A(t)$. The droplet is assumed to be small compared to the wavelength in water $\lambda_0$ and the acoustic pressure can be considered homogeneous around the droplet. Integration of the the momentum equation in water then gives:
\begin{equation}
	\label{eq:p_far}
	p(R_{out},t)= (p_{atm}+p_A(t)) + \rho_0 \left(R\ddot{R}+\frac{3}{2}\dot{R}^2\right),
\end{equation}
with $p(R_{out},t)$ the pressure on the outside of the droplet interface. $p_{atm}$ is the atmospheric pressure, $R(t)$ is the droplet radius, and its overdots represent the interface velocity and acceleration, respectively. The pressure jump across the interface is expressed using the normal stress balance:
\begin{equation}
\label{eq:p_interface}
	p(R_{in},t)-p(R_{out},t) = 4(\mu_0-\mu_1)\frac{\dot{R}}{R}+\frac{2\sigma}{R}, 
\end{equation}
with $\mu_i$ the viscosity of medium $i$. While in the analogous derivation of the Rayleigh-Plesset equation for bubbles the equation is closed by the highly compressible and uniform gas pressure~\cite{Leighton1994}, here we need to evaluate the complete acoustic pressure distribution within the droplet. 
A classical acoustic derivation of the particle velocity $v$ leads to the well-known spherical Bessel equation:
\begin{equation}
\label{eq:Bessel}
 	x^2 \frac{\partial^2 v}{\partial x^2} + 2 x \frac{\partial v}{\partial x} + (x^2 -2) v = 0~.
\end{equation}
Here, $x=k_1r$, with $k_1=\omega/c_1$ the wavenumber, $c_1$ the speed of sound in the droplet, and $r$ the radial coordinate. Avoiding the unphysical divergence at $r=0$, the solution to Eq.~(\ref{eq:Bessel}) is a spherical Bessel function of the first kind:
\begin{equation}
	v(r,t) = f(t)j_1(x)=f(t){\frac {\sin x - x\cos x}{x^{2}}}.
\end{equation}
Writing $X= k_1R$, the boundary condition at the droplet interface $v(R,t) = \dot{R}$ yields:
\begin{equation}
\label{eq:velocity}
	v(r,t)= \dot{R}~\frac{j_1(x)}{ j_1(X)}.
\end{equation}  
The pressure distribution is then found by inserting Eq.~(\ref{eq:velocity}) in the mass conservation and compressibility equation and using the Bessel function recurrence relation $j_1'(x)=j_0(x)-2j_1(x)/x$~:
\begin{equation}
\label{eq:pressvar}
	\frac{\partial p}{\partial t} = -\dfrac{1}{\beta_1} \left(\frac{\partial v}{\partial r} + 2\dfrac{v}{r} \right)  = - \rho_1 k_1 c_1^2\dot{R} \dfrac{j_0(x)}{j_1(X)},
\end{equation}
where $\beta_1 = 1/\rho_1c_1^2$ is the compressibility of the droplet. Note that, owing to the $j_0(x) =sinx/x$ sinc term, the pressure amplitude will always be maximum in the center of the droplet. Integrating Eq.~(\ref{eq:pressvar}) for small oscillation amplitudes ($R = R_0(1+\epsilon$), $\epsilon \ll 1$) with $R_0$ the resting radius and $X_0=k_1R_0$, with the initial boundary condition $p(r,0) = p_{atm} + 2\sigma/R_0$ at $t=0$ and evaluating in $r = R$ yields: 
\begin{equation}
\label{eq:pressure_time}
	p(R_{in},t)=(p_{atm} + 2\sigma/R_0) - \rho_1 k_1 c_1^2 \frac{j_0(X)}{j_1(X_0)} (R-R_0). 
\end{equation}
Note that this approximation only holds when $j_1(X_0)$ is not near its zero-crossing.   
Combining Eqs.~(\ref{eq:p_far}), (\ref{eq:p_interface}) and (\ref{eq:pressure_time}) now gives the droplet dynamics equation:
\begin{equation}
\label{eq:pressure_diff_eq4}
\begin{split}
	\rho_0 \left(R\ddot{R}+\dfrac{3}{2}\dot{R}^2\right) = -\rho_1 k_1 c_1^2\frac{j_0(X)}{j_1(X_0)} R\left(1-\frac{R_0}{R}+\frac{\dot{R}}{c_0}\right)\\ 
	- 4(\mu_0-\mu_1)\frac{\dot{R}}{R}  - 2 \sigma \left(\frac{1}{R}-\frac{1}{R_0}\right) - p_A(t)~.   
\end{split}
\end{equation}
{The reradiated pressure scattered by the droplet induces a compression of the surrounding medium. The effect of acoustic reradiation on the droplet dynamics can be expressed by an additional pressure term  $\frac{R}{c_0}\frac{\partial p}{\partial t}$ at $r=R$~\cite{Prosperetti1986}}, effectively adding a damping term to the set of equations. The above equations can also be extended to include the acoustic interaction with a rigid wall through the addition of a pressure term $\rho_0 \frac{\partial}{\partial t} \left(\frac{R^2\dot{R}}{2d}\right)$ representing the reflected scattering, with $d$ the distance to the wall~\cite{Marmottant2006}. For a droplet at the wall, $d=R$, and the left-hand side of Eq.~\ref{eq:pressure_diff_eq4} takes t\textit{}he form $\rho_0 \left(\frac{3}{2} R\ddot{R}+2\dot{R}^2\right)$, see SI.2.   
Equation~(\ref{eq:pressure_diff_eq4}) can be solved numerically to obtain the resonance behavior of the system. The pressure in the center of the droplet can be obtained by re-evaluating Eq.~(\ref{eq:pressvar}) in $r = 0$ instead of $r = R$~:
\begin{equation}
\label{eq:pressure_origin}
	p_{drop}=(p_{atm} + 2\sigma/R_0) - \rho_1 k_1 c_1^2 \frac{(R-R_0)}{j_1(X_0)}~. 
\end{equation}

\begin{figure}[t]
\includegraphics[width=0.8\columnwidth]{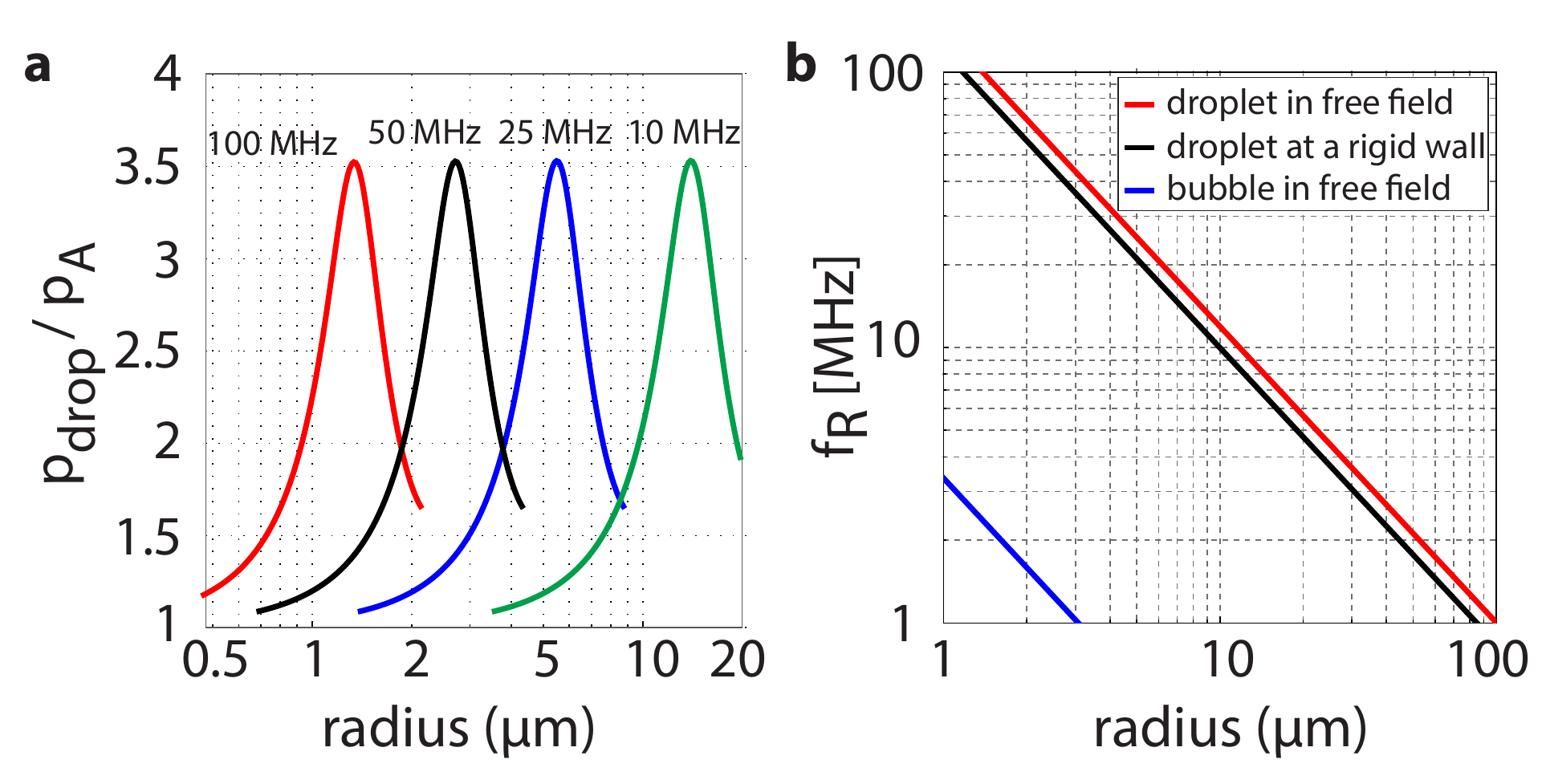}
\caption{(a) Calculated theoretical resonance curves for different driving frequencies. (b) Calculated resonance frequency $f_R$ as a function of size: droplet in free-field (red), droplet at a rigid wall (black) and bubble in free-field (blue).}
\label{fig:1}
\end{figure}

Linearization of Eq.~(\ref{eq:pressure_diff_eq4}) gives a relation between the angular eigenfrequency of the droplet $\omega_0$ and its resting radius: 
\begin{equation} 
\label{eq:resfreq}
	\rho_0(\omega_0R_0)^2 = \dfrac{\rho_1\omega_0 R_0 c_1}{\frac{c_1}{\omega_0R_0}- cot(\frac{\omega_0R_0}{c_1})} + \frac{2\sigma}{R_0}. 
\end{equation}
Using a first-order expansion $cot(z) \simeq 1/z-z/3+\mathcal{O}(z^3)$ and {by neglecting the interfacial tension term}, Eq.~(\ref{eq:resfreq}) reduces to a simple classical form:
\begin{equation}
\label{eq:eigenfreq}
	f_0 \simeq \frac{1}{2 \pi R_0}\sqrt{{\dfrac{3\rho_1c_1^2}{\rho_0}}}~,
\end{equation}
with $f_0 = {\omega_0}/{2\pi}$ the eigenfrequency of the droplet.
The same approach leads to an expression for the damping of the system in canonical form:
\begin{equation} \label{eq:damping}
	\delta = \frac{\omega_0 R_0}{2c_0}+\frac{2(\mu_0-\mu_1)}{\rho_0\omega_0R_0^2}. 
\end{equation}
As a result of the negligible contribution of viscous damping, {which is three orders of magnitude smaller than the first term in Eq.~(\ref{eq:damping})},  this then further reduces to:  
\begin{equation}
\label{eq:damping_simple}
	\delta \simeq \frac{c_1}{c_0} \sqrt{\frac{3 \rho_1}{4\rho_0}}~.
\end{equation}
For a droplet at the wall the linearization simply adds a prefactor $\sqrt{2/3}$ to both Eqs.~(\ref{eq:eigenfreq}) and (\ref{eq:damping_simple}). Equation~(\ref{eq:damping_simple}) also shows that the resonance arises from a speed of sound mismatch between the droplet and its surrounding medium, i.e. when the speed of sound ratio goes to 1, the system becomes overdamped, without a resonance effect, while the largest speed of sound mismatch produces the strongest resonance. 

The resonance curves, solutions of Eq.~(\ref{eq:pressure_diff_eq4}) and computing the maximum pressure in the drop using Eq.~(\ref{eq:pressure_origin}) for frequencies of 10, 25, 50 and 100~MHz were calculated using the {\tt ODE45} solver in Matlab and are displayed in Fig.~\ref{fig:1}(a). The physical parameters for PFP and water that were used for the calculations are listed in SI.3 and were extracted from \cite{NIST,Morgado2011,Kandadai2010}. The resonance frequency $f_R$ is plotted against the droplet size in Fig.~\ref{fig:1}(b) for a droplet in free-field and for a droplet against a rigid wall. Two direct results from these plots are that (1), unlike superharmonic focusing, the acoustic resonance strength has very little dependency on droplet size (see also SI.4) and (2), the resonance is expected to have a high quality factor and a frequency about 50 times higher than that of free gas bubbles, see the corresponding Minnaert bubble resonance frequency $f_{M} =3.3~\textup{$\mu$mMHz}/R_0 $ plotted in Fig.~\ref{fig:1}(b)~\cite{Leighton1994}.

\begin{figure}[t]
\includegraphics[width=0.7\columnwidth]{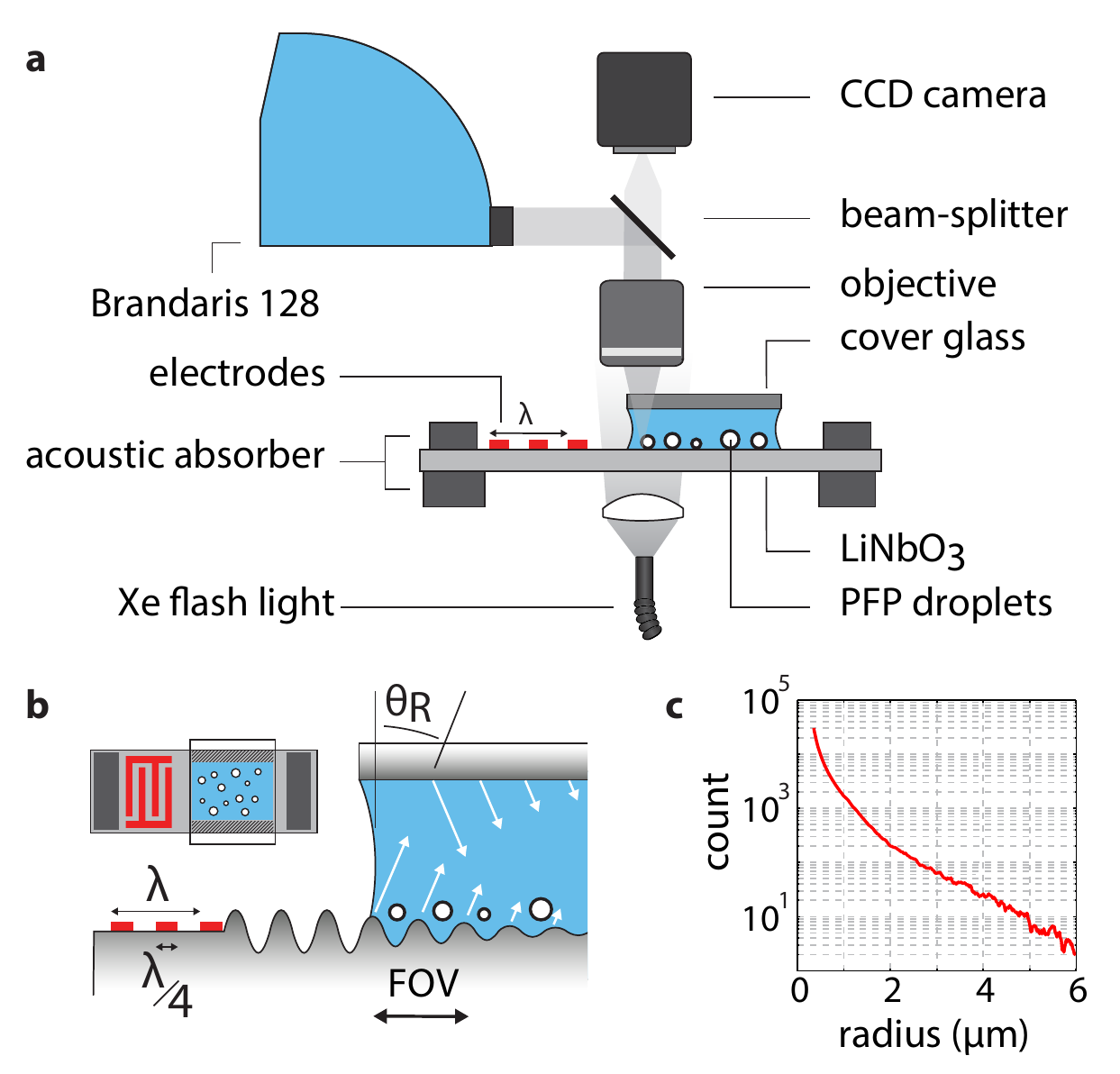}
\caption{(a) Schematic of the experimental setup. The vaporization of PFP droplets is imaged using the Brandaris 128 ultra-high speed camera. (b) A SAW device generates 	a longitudinal pressure wave in the fluid at the Rayleigh angle $\theta_R$. (c) Typical size distribution of the droplets used in this study.}
\label{fig:2}
\end{figure}

Owing to the high eigenfrequencies expected, the response of PFP droplets was measured experimentally at frequencies of 19.6 and 45.4~MHz. The droplets {were resting on the piezoelectric substrate} on which surface acoustic waves (SAW) were generated using an interdigitized transducer (IDT), see Fig.~\ref{fig:2}(a). In contact with water, the SAW generates a longitudinal bulk acoustic wave at the Rayleigh angle $\theta_R$~$\approx$ 23\degrees~\cite{Shilton2008, Wang2011b, Yeo2014}, see Fig.~\ref{fig:2}(b). The use of a SAW device prevents any nonlinear propagation in the bulk of the medium, leaving a purely sinusoidal excitation.
Straight electrode IDTs with a single aluminum electrode pair per wavelength (60 pairs, thickness of 750~nm, aperture of 1~cm) were fabricated on a 128\degrees\ rotated Y-cut X-propagating lithium niobate (LiNbO${_3}$, Roditi, United Kingdom) wafer using standard soft lithography techniques. The IDTs were actuated by a 50-cycle sinusoidal ultrasound pulse generated by a waveform generator (model 8026, Tabor Electronics) connected to a  50~dB linear power amplifier (350L, E\&I). A sound absorbing silicone rubber (PDMS, Dow Corning) was placed both below and above the end of the piezoelectric substrate to reduce acoustic reflections. 

A perfluoropentane (PFP) droplet emulsion was prepared as in~\cite{Reznik2012}. Its size distribution was measured using a Coulter counter, see Fig.~\ref{fig:2}(c). The suspension was loaded by capillary suction in a chamber that was approximately 100~$\mu$m in height. The chamber was located directly above the piezoelectric substrate, open at the front end to allow for a direct coupling between the SAW and the liquid, and closed above using a microscope cover slip {(24~mm length)}, see details in Fig.~\ref{fig:2}(b). 

\begin{figure}[t]
	\includegraphics[width=0.8\columnwidth]{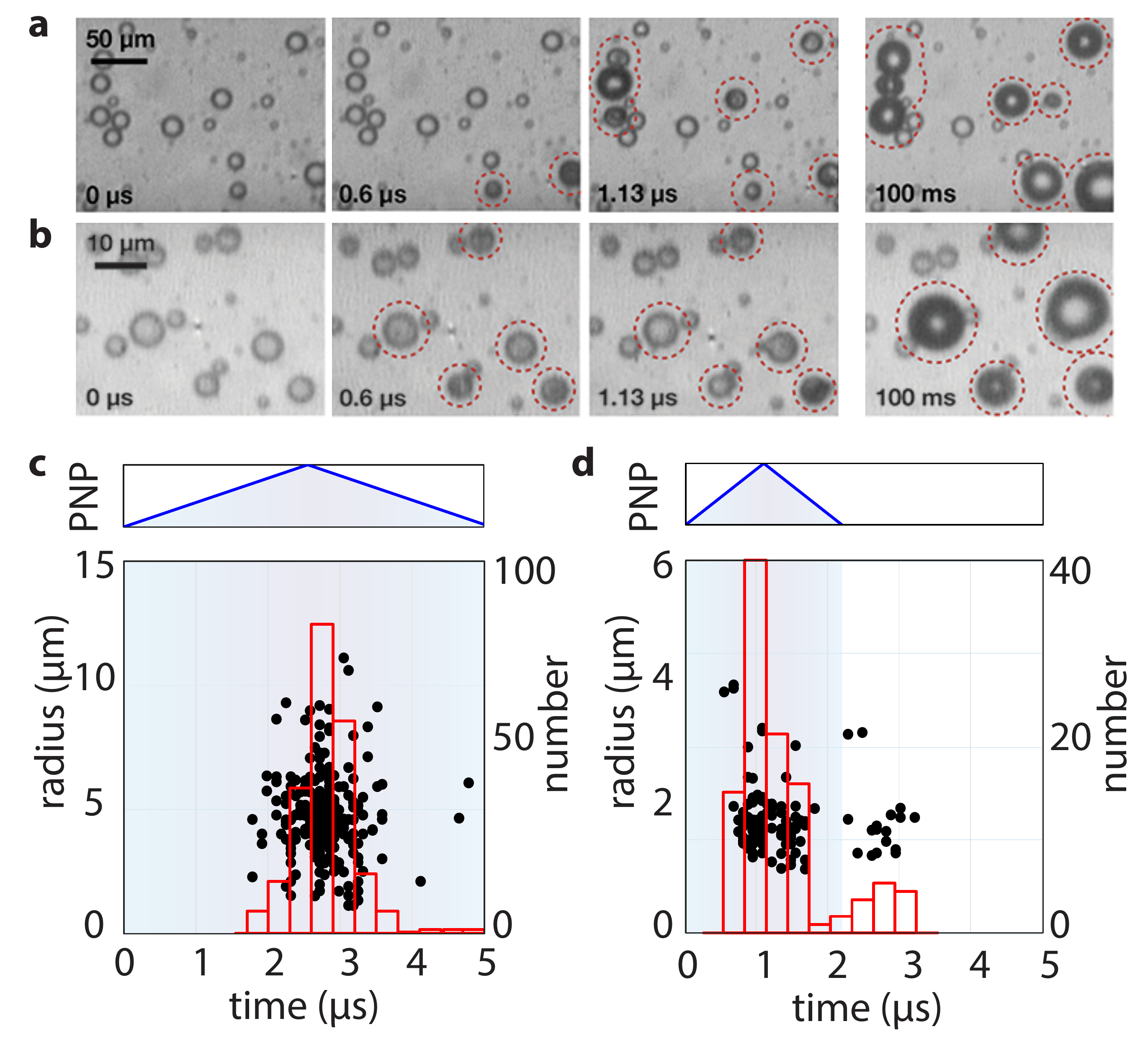}
	\caption{Vaporization of PFP droplets driven at 19.6~MHz (a) and 45.4~MHz (b) imaged at 15~Mfps. The nucleated droplets are marked by the red dotted circles. Stable 		bubbles are formed from all nucleated droplets as can be observed from the image captured 100~ms later. Number of nucleation events over time and the corresponding 		radius of the droplet at a driving frequency of 19.6~MHz (c) and 45.4~MHz (d).}
	\label{fig:3}
\end{figure}

Droplet vaporization was imaged using an inverted microscope (Olympus BX-FM) equipped with a 20$\times$ magnification objective (Olympus SLMPlan N) coupled to the Brandaris 128 ultra high-speed camera~\cite{Chin2003, Gelderblom2012} operated at 15~million frames per second (Mfps) to record the time and location of droplet nucleation. The imaging resolution was 0.29~$\mu$m per pixel. The field of view (FOV) was positioned close to the meniscus of the liquid, at the front end of the chamber, to minimize interference caused by acoustic reflections from the top of the chamber. Three successive high-speed recordings of 128~frames each were acquired at an interval of 100~ms. Droplets were vaporized during the second recording and the first frames of the third recording were used to image the bubbles formed. The high-speed imaging frames were processed with an automated image analysis procedure programmed in Matlab (The MathWorks, Natick, MA). All experiments were performed at 20\degrees C.

Figure~\ref{fig:3} shows an image sequence of the vaporization of PFP droplets driven at frequencies of 19.6~MHz (a) and 45.4~MHz (b). {The droplets that underwent nucleation} are marked by the red-dotted circles. At 19.6~MHz, the nucleated bubbles grow rapidly due to rectified heat transfer under acoustic forcing~\cite{Shpak2013}. The bubble size subsequently decreased within microseconds after the ultrasound driving was stopped. ADV at 45.4~MHz is less violent than at 19.6~MHz. In particular, the bubbles that nucleated inside the droplets were much smaller at 45.4~MHz. Note that every nucleation site produced a stable bubble at both driving frequencies, as can be observed from the images captured after 100~ms, i.e. none of the nucleated bubbles were observed to recondense~\cite{Shpak2013a}.

Figures~\ref{fig:3}(c,d) show a total of 217 individual nucleation events for a driving frequency of 19.6~MHz and 120 individual nucleation events for a frequency of 45.4~MHz, represented as the droplet radius versus the time at which nucleation occurs. The number of droplets that nucleated is shown in the red histogram. The normalized PNP of the driving pulse that results from the superposition of waves transmitted by the 60 electrode pairs is shown in the top panels of Figs.~\ref{fig:3}(c,d). {Since the electrode pairs are spaced by 1 wavelength, the typical surface wave has a triangular shape originating from 50~superimposed waves. The pressure profile was experimentally verified by optical hydrophone recordings with one end of the LiNbO$_3$ substrate submerged in water.} The bubbles are thus more likely to nucleate at the maximum PNP, namely at the end of the 50~cycle IDT driving pulse. {Note the presence of a second event of vaporizations in Fig.~\ref{fig:3}(d) where a small number of droplets nucleated after the primary wave had passed, most probably a result of an internal reflection in the IDT device}. Also, the absence of structure in the data presented in Figs.~\ref{fig:3}(c,d) suggests that there is no size dependency on the timing of the vaporization.  

\begin{figure}[t]
\includegraphics[width=0.71\columnwidth]{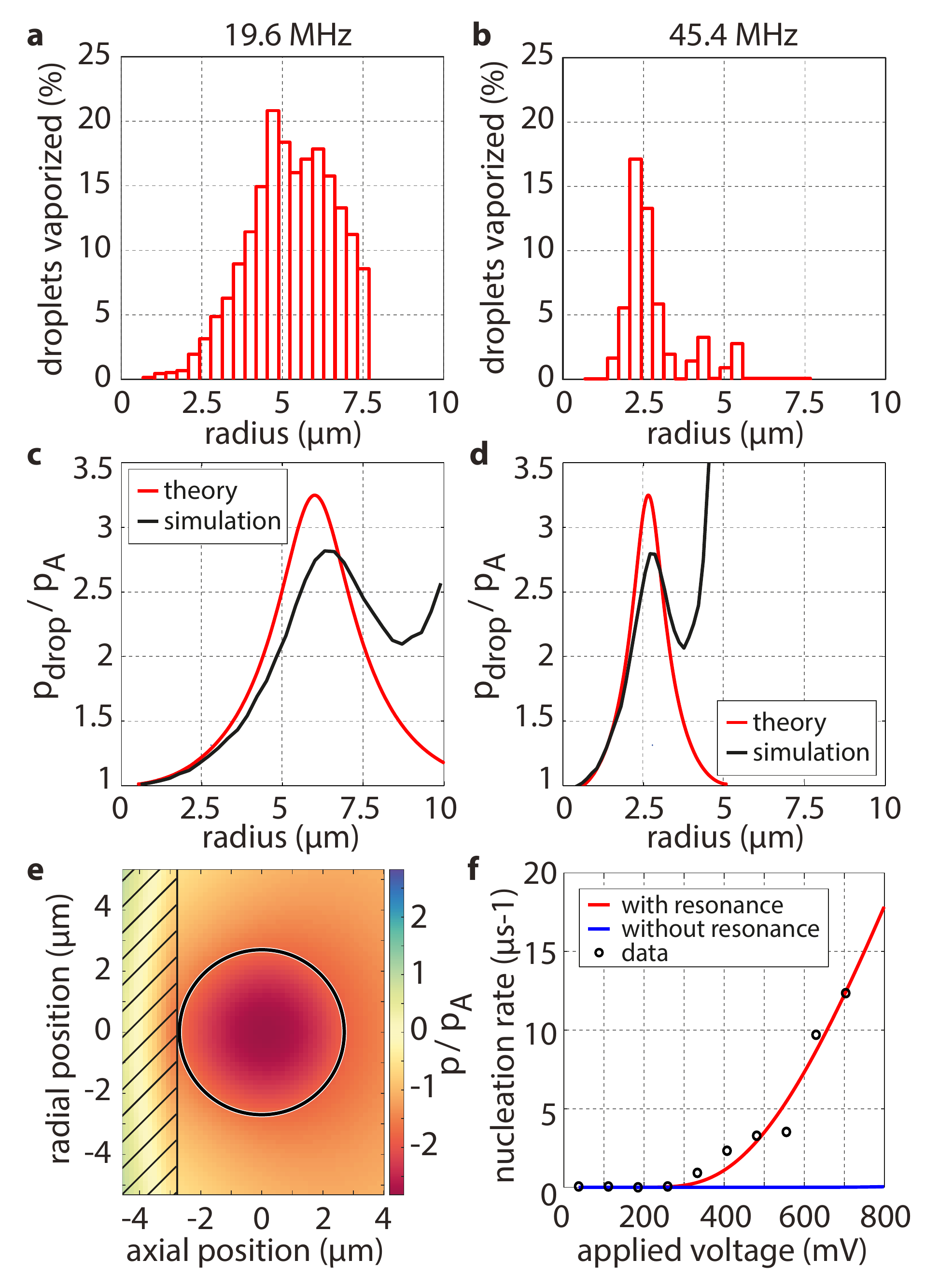}
\caption{Resonant vaporization detected at a frequency of 19.6~MHz (a) and 45.4~MHz (b). Calculated and simulated size-dependent pressure amplification factor within 		the droplet at a frequency of 19.6~MHz (c) and 45.4~MHz (d). (e) Snapshot of the simulated pressure field for a 2.7~$\mu$m radius droplet resonant at a frequency of 45.4~MHz. (f) 	Droplet activation probability with and without the resonance effect {where 'data' represents the experimental droplet activation probability.}}
\label{fig:4}
\end{figure}

The proportion of droplets activated by the ultrasound wave with respect to the total number of droplets present {in each separate bin is shown in the histograms} displayed in Figs.~\ref{fig:4}(a,b) for the driving frequencies of 19.6~MHz and 45.4~MHz, respectively. The resonance curves calculated from the proposed theory are displayed in Figs.~\ref{fig:4}(c,d) (red lines). The theoretical resonance peaks in Fig.~\ref{fig:4}(c,d) are corrected for the presence of a rigid wall and closely match the experimental peaks in Figs.~\ref{fig:4}(a,b). 

Numerical simulations were performed on the basis of the geometry of Fig.~\ref{fig:2}(b) to provide further insight in the resonance behavior. The simulations were axisymmetric and computed on a GPU using k-wave, an open source Matlab toolbox for time domain ultrasound simulations in complex media~\cite{Treeby2010a}, see SI.5. The grid size was 0.30~$\mu$m at a frequency of 19.6~MHz and 0.15~$\mu$m at a frequency of 45.4~MHz. The results are plotted in Figs.~\ref{fig:4}(c,d)(black lines). The resulting pressure field in and around the droplet is displayed in the snapshot of Fig.~\ref{fig:4}(e), taken at $t=375$~ns, i.e after 17~cycles of ultrasound, see also Supplementary Video. The amplification factor of the first resonance, as well as its location, is in very good agreement with the theoretical model, although the simulated amplification is $\sim 15 \% $ lower than predicted by the theory. Interestingly, the presence of a second mode (see SI.5) is visible at 45.4~MHz. It appears that the experimental data for a driving frequency of 45.4 MHz (Fig.~\ref{fig:4}(b)) indeed displays a secondary nucleation peak around a radius of 5~$\mu$m. This higher mode appears to have a smaller effect on the ADV threshold despite the large pressure amplification (factor 10) found in simulation. This is most likely a limitation of the present rigid numerical simulation since higher order non-axisymmetric modes can induce droplet deformation, which may have a significant impact on the pressure distribution within the droplet.

The statistical increase in the number of vaporization events as a function of the negative acoustic pressure amplitude was investigated by varying the amplitude of the 50-cycle ultrasound pulses at a frequency of 19.6~MHz. {The envelop of the pressure wave, depicted in the top panels of Fig.~\ref{fig:3}c and d, combined with the measured timing of the event, were used to determine the voltage at which each droplet vaporized.} The resulting nucleation rate is displayed in Fig.~\ref{fig:4}(f) (open circles).
Classical nucleation theory \cite{Caupin2006, Karthika2016} dictates that the nucleation rate is a function of the surface tension of the liquid $\sigma_{\textup{PFP}} \simeq 10$~mN/m~\cite{Kandadai2010}, of the ambient temperature $T_{amb} = 293$ K and of the ambient pressure:
\begin{equation} \label{eq:nuc_rate}
\Gamma 	\propto e^{\left(  -\dfrac{16\pi}{3k_BT_{amb}} \dfrac{\sigma_{\textup{PFP}}^3}{\left( p_v - p_{atm} - p_{drop}cos(\omega t)  \right) ^2}   \right)}.
\end{equation}
Here, $p_v$ is the vapor pressure of the liquid and $k_B$ is the Boltzmann constant. {The vapor pressure of PFP can be estimated using Antoine's law and ranges from 65~kPa at a temperature of 15$^\circ$C to 95~kPa at 25$^\circ$C.} As a result, and considering the large peak negative pressures typically required to induce cavitation, $\left( p_v - p_{atm}\right)/p_{drop} \ll 1$. With this simplification, Eq.~(\ref{eq:nuc_rate}) can be {volume-}integrated during the negative pressure phase of the ultrasound cycle to determine the average nucleation rate:
\begin{equation} \label{eq:av_nuc_rate}
	<\Gamma> ~\propto \int^{R_{0}}_{0} r^2 \left( 1- erf\left(\dfrac{1}{p(r)} \sqrt{\dfrac{16\pi\sigma_{\textup{PFP}}^3}{3k_BT_{amb}}} \right)\right) dr,
\end{equation}
see details provided in SI.6.
At resonance the pressure amplification factor due to the resonance effect $p_{drop}/p_{A}\simeq 3$, see Figs.~\ref{fig:4}(c,d). When the SAW devices are driven within their linear range, the acoustic pressure is proportional to the driving voltage. Since the droplet concentration is contained in the prefactor, this expression can be used to fit the experimental data and provide an estimate of the local acoustic driving pressure, which is the only free parameter in Eq.~(\ref{eq:av_nuc_rate}). {The least-squares fit is shown in Fig.~\ref{fig:4}(f) and gives $p_A/V \simeq$~40~MPa/V, leading to an acoustic pressure at the droplet location of $\sim $ 20~MPa for an excitation voltage of 500 mV}. {Note that these numbers only constitute a rough estimate since, on the one hand, classical nucleation theory is known to overestimate the peak negative pressures required for cavitation~\cite{Caupin2006} and, on the other hand, the range of data available for fitting is limited.} It should also be noted that this pressure is of the same order as those typically used for ADV. Interestingly, nucleation theory as presented in Eq.~(\ref{eq:av_nuc_rate}) allows to predict the importance of the resonance effect by calculating the average nucleation rate without resonance, namely for $p_{drop}/p_{A}=1$ (with the same prefactor), see Fig.~\ref{fig:4}(f) (blue line). Thus, it is clear that a pressure amplification factor of 3 has a dramatic effect on droplet vaporization behavior.

In conclusion, it was shown that efficient vaporization can be achieved by driving phase-change PFP droplets at their fundamental resonance frequency. Good agreement was found between the modeled size-dependent pressure amplification within the droplet and the measured size dependent vaporization probability. Resonance-induced vaporization is a new phenomenon, with important impact on potential ADV strategies and on the understanding on previous experimental observations. In addition, this work shows the potential of using monodisperse phase-change agents driven by a matching resonance frequency to boost the efficiency of ADV. 

We thank James Friend for stimulating discussions. This work was funded by NanoNextNL, a micro and nanotechnology consortium of the Government of the Netherlands and 130 partners. 

\bibliography{Bibliography_5}
\onecolumngrid
\clearpage
\end{document}


\title{Supplementary Information\\
High-frequency acoustic droplet vaporization\\ is initiated by resonance}
\author{Guillaume Lajoinie\footnotemark}
\affiliation{Physics of Fluids Group, MESA+ Institute for Nanotechnology, Technical Medical (TechMed) Center, University of Twente, P.O. Box 217, 7500 AE Enschede, The Netherlands}       
\author{Tim Segers\footnotemark[\value{footnote}] \footnotetext{Both authors contributed equally to this manuscript}}
\affiliation{Physics of Fluids Group, MESA+ Institute for Nanotechnology, Technical Medical (TechMed) Center, University of Twente, P.O. Box 217, 7500 AE Enschede, The Netherlands}       
\author{Michel Versluis}
\affiliation{Physics of Fluids Group, MESA+ Institute for Nanotechnology, Technical Medical (TechMed) Center, University of Twente, P.O. Box 217, 7500 AE Enschede, The Netherlands}       

\maketitle

\section{SI.1 Derivation of the droplet dynamics equation}
\subsection{The problem}
Assume a spherical droplet of PFC in medium 0 (density $\rho_0$, speed of sound $c_0$). The radius of the PFC droplet is $R$ and its density is $\rho_1$. The PFC has a speed of sound~$c_1$. The droplet is driven by an external pressure field $p_A(t)=p_A sin(\omega t)$, with the angular frequency $\omega = 2\pi f$ and $f$ the frequency of the driving pulse. We would like to calculate the pressure distribution in the droplet $p(r,t)$ as a result of this external pressure field $p_A(t)$.

\begin{figure}[h]
\begin{center}
\begin{tikzpicture}
	\draw[step=0.5cm,gray,very thin] (-2.9, -2.9) grid (2.9,2.9);
	\filldraw[cyan!10!white, draw=black] (0,0) circle (2cm) ;
	\filldraw (0,0) circle (2pt) ; 
	\draw[gray,thin,dashed] (-2,0) -- (2,0);
	\draw[gray,thin,dashed] (0,-2) -- (0,2);
	\draw[thick,->] (0,0) -- (135:2cm); 
	\node at (-0.3, 1.0) {$R(t)$} ;
	\draw [thick,-stealth] (-3.5,2.5) -- (-2.5,2.5);
	\draw [thick,-stealth] (-3.5,2) -- (-2.5,2);
    \draw [thick,-stealth] (-3.5,1.5) -- (-2.5,1.5);
    \draw [thick,-stealth] (-3.5,1) -- (-2.5,1);
    \draw [thick,-stealth] (-3.5,0.5) -- (-2.5,0.5);
    \draw [thick,-stealth] (-3.5,0) -- (-2.5,0);
    \draw [thick,-stealth] (-3.5,-0.5) -- (-2.5,-0.5);
    \draw [thick,-stealth] (-3.5,-1) -- (-2.5,-1);
    \draw [thick,-stealth] (-3.5,-1.5) -- (-2.5,-1.5);
    \draw [thick,-stealth] (-3.5,-2) -- (-2.5,-2);
    \draw [thick,-stealth] (-3.5,-2.5) -- (-2.5,-2.5);
    
	\node at (1,-0.8) {$p(R_{in},t)$};
    \node at (2.9,-2.0) {$p(R_{out},t)$};
	\draw [-] (0.4,-1.1) to [bend right] (1.4142,-1.4142) ;
	\draw [-] (2.2,-1.8) to [bend right] (1.4142,-1.4142) ;
	\node at (1.5,-3) {$p_{atm}$};
    \node at (-3,-3) {$p_A(t)$};
    
    \filldraw (1.4142,-1.4142) circle (1pt) ;
	\node at (2.0,2.3) {$\rho_0, \mu_0$};
	\node at (2.0,1.8) {$c_0, v_0$};
	\node at (-0.8,-0.8) {$\rho_1, \mu_1$};
	\node at (-0.8,-1.3) {$c_1, v_1$};
	
\end{tikzpicture}
\end{center}
\caption{Schematic of the droplet system with the relevant physical parameters. A plane wave enters from the left.} 
\label{fig:SI1}
\end{figure}

\subsection{The solution}
The approach is to connect the pressure in the droplet $p(r,t)$ to the acoustic driving pressure $p_A(t)$ in medium 0 as a sum of pressure differences: inside the droplet, at the interface and in the surrounding medium: 
\begin{equation}
p(r,t)=\Delta p_{droplet}+\Delta p_{interface}+\Delta p_{0}+p_{atm}+p_A~,
\end{equation}
which originates from 
\begin{equation}
p(r,t) = p(r,t) - p(R_{in},t) +  p(R_{in},t) -p(R_{out},t) + p(R_{out},t) - (p_{atm}+p_A) + (p_{atm}+p_A)~,
\end {equation}
with $p_{atm}$ the atmospheric pressure.
Here we derive the equations for each of these pressure differences. Moreover, we start by finding the wave equation in the droplet for $v(r,t)$ instead of the more classical approach of solving the pressure wave equation. This circumvents the potential problem of a discontinuous pressure jump at the droplet interface.   

\subsection*{Assumptions}
For a frequency of 20 MHz, the wavelength of the driving pulse in water is 75 $\mu$m, which is 15 times larger than the resonant size of the PFC droplets (to be shown later). Therefore we assume that there is no phase relationship, or direction of the wave, along the interface of the droplet. We therefore also assume that the wave in the droplet has radial symmetry, i.e. there is no dependency on the angles $\vartheta$ and $\varphi$. Furthermore, we follow the acoustic approximation such that the nonlinear hydrodynamic equations lead to the linear wave equation for the ultrasound wave in the droplet: $\rho = \rho_1 + \rho'$ with $\rho' \ll \rho_1$ and $p = p_1 + p' $ with $p' \ll p_1$.
    
\subsection{Basic equations}
The system can be solved by using three basic equations: definition of the compressibility, the momentum equation, and continuity.
\subsection{a. Compressibility}
The compressibility of the droplet medium with volume $V$ and pressure $p$ is given in differential form:
\begin{equation}
\label{eq:compress}
\beta=-\frac{1}{V}\frac{\partial V}{\partial p}=\frac{1}{\rho_1}\frac{\partial \rho}{\partial p}~.
\end{equation}
The change in pressure and density ${\partial \rho}/{\partial p}$ is related to the speed of sound:
\begin{equation}
\frac{\partial \rho}{\partial p} = \frac{1}{c^2} \rm = \beta \rho ~ .
\end{equation}
In linear acoustics, the perturbation is carried out to the first order, and only the first-order terms are kept in the governing equations. A Taylor series expansion then gives:
\begin{align}
\rho - \rho_1 & = \left(\frac{\partial \rho}{\partial p}\right) (p - p_1)~, \\
\rho' & =\frac{1}{c^2} p'~,
\end{align} 
where $\rho '$ and $p'$ indicate the first order density and pressure \emph{fluctuations} linearized at the linearization points $\rho_1$ and $p_1$.
In the remainder we will write $\rho$ and $p$ where we actually refer to $\rho '$ and $p'$. The constitutive equation, Eq.~(\ref{eq:compress}), then simplifies to:
\begin{equation}
\label{eq:compress:lin}
\rho=\beta_1 \rho_1 p~, 
\end{equation} 
with 
\begin{equation}
\label{eq:sos:lin}
\beta_1=\frac{1}{\rho_1 c_1^2}~. 
\end{equation}

\subsection{b. Continuity}
The continuity equation for the droplet system is given in differential form:
\begin{equation}
\frac{\partial \rho}{\partial t} + \nabla \cdot (\rho \bm{v}) = 0~,
\end{equation}
which in spherical coordinates with radial symmetry in linearized form reduces to:
\begin{equation}
\frac{\partial \rho}{\partial t} = - \rho_1 \frac{1}{r^2} \frac{\partial}{\partial r} (r^2v)
= - \rho_1 \frac{\partial v}{\partial r} - \rho_1 \frac{2v}{r}~.
\end{equation}
With Eq.~(\ref{eq:compress:lin}):
\begin{equation}
\label{eq:cont:lin}
\beta_1 \frac{\partial p}{\partial t} = - \frac{\partial v}{\partial r} - \frac{2v}{r}~.
\end{equation}

\subsection{b. Momentum}
The momentum equation for the droplet system is given in differential form:
\begin{equation}
\rho \frac{\partial \bm{v}}{\partial t} + \rho \bm{v} \cdot \nabla \bm{v}  = -\nabla p,
\end{equation}
which in linearized form reduces to:
\begin{equation}
\rho_1 \frac{\partial v}{\partial t} + \frac{1}{2} \rho_1 \frac{\partial{v^2}}{\partial r}  = -\frac{\partial p}{\partial r},
\end{equation}
and neglecting the non-linear term  ${\partial{v^2}}/{\partial r}$ we get:
\begin{equation}
\label{eq:momentum:lin}
\frac{\partial p}{\partial r} = -\rho_1 \frac{\partial v}{\partial t}~.  
\end{equation}

\subsection{Acoustic wave equation}
We now take the spatial and time derivatives of Eqs.~(\ref{eq:cont:lin}) and~(\ref{eq:momentum:lin}), respectively :
\begin{align}
\beta_1 \frac{\partial^2 p}{\partial r \partial t} & = - \frac{\partial^2 v}{\partial r^2} - \frac{2}{r}\frac{\partial v}{\partial r} + \frac{2v}{r^2}~, \\
\beta_1 \frac{\partial^2 p}{\partial r \partial t} & = - \beta_1 \rho_1 \frac{\partial^2 v}{\partial t^2}~.  
\end{align}
By equalizing these two equations we get:
\begin{equation}
 - \frac{\partial^2 v}{\partial r^2} - \frac{2}{r}\frac{\partial v}{\partial r} + \frac{2v}{r^2} = - \beta_1 \rho_1 \frac{\partial^2 v}{\partial t^2}~,
\end{equation}
which can be rewritten to:
\begin{equation}
 r^2 \frac{\partial^2 v}{\partial r^2} + 2 r \frac{\partial v}{\partial r} + \left[(k_1 r)^2 -2\right] v = 0~,
\end{equation}  
by substituting the wavenumber $k_1=\omega/c_1$ with $c_1^2 = 1/\beta_1 \rho_1$, Eq.~(\ref{eq:sos:lin}), and $\partial^2 v / \partial t^2 = -\omega^2 v$ for a sinusoidal finite amplitude wave $v=v(r)sin\omega t$.  

\subsection{Velocity field} 
Rewriting with $x=k_1r$ gives a spherical Bessel function of order~1:
\begin{equation}
 x^2 \frac{\partial^2 v}{\partial x^2} + 2 x \frac{\partial v}{\partial x} + (x^2 -2) v = 0~,
\end{equation}
and the solution is:
\begin{equation}
\label{eq:x_solution}
j_{1}(x)={\frac {\sin x}{x^{2}}}-{\frac {\cos x}{x}} = \frac {\sin x - x\cos x}{x^2}~. 
\end{equation}
The solution of the spherical wave equation for the velocity in the droplet is given by Eq.~\ref{eq:x_solution} and by taking into account the time-dependency of the wave:
\begin{equation}
\label{eq:x_t_solution}
v(r,t)=f(t)j_1(x)~.
\end{equation}
The boundary condition at the droplet interface $v(R,t) = \dot{R}$ gives:
\begin{equation}
\label{eq:boundcond}
v(R,t)=\dot{R}=f(t)j_1(X)~,
\end{equation}
with $X= kR$ and hence:
\begin{equation}
\label{eq:velocity}
v(r,t)=\dot{R}~\frac{j_1(x)} {j_1(X)}.
\end{equation}

\subsection{Pressure in the droplet - medium 1}
The pressure distribution in the droplet can now be found by introducing the velocity, Eq.~\ref{eq:velocity}, in the continuity equation, Eq.~\ref{eq:cont:lin}:
\begin{equation}
 \frac{\partial p}{\partial t} = - \rho_1 c_1^2 \frac{\dot{R}}{j_1(X)} \left(\frac{\partial}{\partial r}~j_1(x) + \frac{2j_1(x)}{r}\right)
\end{equation}
and with Eq.~\ref{eq:x_solution}:
\begin{equation}
\begin{split}
\frac{\partial}{\partial r}~j_1(x) &=k_1\frac{\partial}{\partial x}\left(\frac {\sin x - x\cos x}{x^2}\right)\\
&=k_1\left(\frac{(cosx-cosx+xsinx)x^2-2x(sinx-xcosx)}{x^4}\right)\\
&=k_1\left(\frac{sinx}{x}-\frac{2(sinx-xcosx)}{x^3}\right)\\
&= k_1j_0(x)-\frac{2k_1j_1(x)}{x},
\end{split}
\end{equation}
where $j_0(x)=sinx/x$ and we can rewrite
\begin{equation}
\label{eq:dpdt_26}
 \frac{\partial p}{\partial t} = - \rho_1 k_1 c_1^2 \dot{R}\frac{j_0(x)}{j_1(X)} ~.
\end{equation}
Note that, owing to the $j_0(x)$ sinc term, the pressure amplitude will always be maximum in the center of the droplet. For small oscillation amplitudes ($R = R_0(1 + \epsilon), \epsilon \ll  1$) with $R_0$ the resting radius and $X_0 = k_1R_0$:
\begin{equation}
\label{eq:partial_pressure_time}
\frac{\partial p}{\partial t} = - \rho_1 k_1 c_1^2  \dot{R}\frac{j_0(x)}{j_1(X_0)}.
\end{equation}
This approximation only holds when $j_1(X_0)$ is not near its zero-crossing.
Integration of Eq.~\ref{eq:partial_pressure_time} then leads to:
\begin{equation}
\label{eq:pressure_time}
p(r,t)-p(r,0)=\int_{0}^{t} \frac{\partial p}{\partial t} dt= - \rho_1 k_1 c_1^2 \frac{j_0(x) }{j_1(X_0)}(R-R_0). 
\end{equation}
Evaluating at the droplet interface ($r=R$) and with the boundary condition $p(r,0) = p_{atm} + 2\sigma/R_0$ at $t=0$ when $R=R_0$ we find:
\begin{equation}
\label{eq:pressure_time2}
	p(R_{in},t)=(p_{atm} + 2\sigma/R_0) - \rho_1 k_1 c_1^2 \frac{j_0(X)}{j_1(X_0)} (R-R_0). 
\end{equation}
Evaluating at the center of the droplet ($r=0$) we find the maximum pressure:
\begin{equation}
\label{eq:pressure_origin}
	p_{drop}=(p_{atm} + 2\sigma/R_0) - \rho_1 k_1 c_1^2 \frac{(R-R_0)}{j_1(X_0)}~. 
\end{equation}
 
\subsection{Pressure at the droplet interface}
The pressure jump across the interface is taken from the balance of the normal stresses:
\begin{equation}
\label{eq:p_interface}
p(R_{in},t)+4\mu_1\frac{\dot{R}}{R}=p(R_{out},t) +\frac{2\sigma}{R}+ 4\mu_0\frac{\dot{R}}{R}~, \\
\end{equation}
with $\mu_i$ the viscosity of medium $i$. Hence:
\begin{equation}
\label{eq:p_interface2}
p(R_{in},t)-p(R_{out},t)= 4(\mu_0-\mu_1)\frac{\dot{R}}{R}+\frac{2\sigma}{R}~.
\end{equation}
 
\subsection{Pressure in the surrounding medium 0}
The pressure at the interface (outside) is given by integration of the momentum equation in medium 0 and is analogous to the derivation of the Rayleigh-Plesset equation, see details in ref.~\cite{Leighton1994}:
\begin{equation}
\label{eq:p_far}
\begin{split}
p(R_{out},t)&=p_{atm}+p_A(t)+ \rho_0 \left(R\ddot{R}+\frac{3}{2}\dot{R}^2\right)~,\\
p(R_{out},t)-(p_{atm}+p_A(t))&= \rho_0 \left(R\ddot{R}+\frac{3}{2}\dot{R}^2\right)~.
\end{split}
\end{equation}

\subsection{Resulting differential equation for the droplet dynamics}
We can now combine Eqs.~\ref{eq:pressure_time2},~\ref{eq:p_interface2} and   and~\ref{eq:p_far} to get to the droplet dynamics equation:
\begin{equation}
- \rho_1 k_1 c_1^2 \frac{j_0(X)}{j_1(X_0)} (R-R_0)=
4(\mu_0-\mu_1)\frac{\dot{R}}{R}+\frac{2\sigma}{R}-\frac{2\sigma}{R_0}+\rho_0 (R\ddot{R}+\frac{3}{2}\dot{R}^2)+p_A(t),
\end{equation} 
or rewritten in the more classical form:
\begin{equation}
\label{eq:pressure_diff_eq}
\rho_0 \left(R\ddot{R}+\frac{3}{2}\dot{R}^2\right)=- \rho_1 k_1 c_1^2 \frac{j_0(X)}{j_1(X_0)} (R-R_0)
-4(\mu_0-\mu_1)\frac{\dot{R}}{R}-2\sigma\left(\frac{1}{R}-\frac{1}{R_0}\right)-p_A(t).
\end{equation} 

\subsection{Acoustic reradiation}
{The reradiated pressure scattered by the droplet induces a compression of the 
surrounding medium. The effect of acoustic reradiation on the droplet can be expressed by an
additional pressure term $\dfrac{R}{c_0}\dfrac{\partial p}{\partial t}$ at $r=R$~\cite{Prosperetti1986}, effectively adding a damping term to the set of equations}. With the help of Eq.~\ref{eq:partial_pressure_time} we get:
\begin{equation}
\frac{R}{c_0}\frac{\partial p}{\partial t}\Big|_{R}=-\rho_1k_1 c_1^2  R\frac{\dot{R}}{c_0}  \frac{j_0(X)}{j_1(X_0)}~. 
\end{equation}

In the full equation it writes as follows: 
\begin{equation}
\label{eq:pressure_diff_eq4}
\rho_0 \left(R\ddot{R}+\frac{3}{2}\dot{R}^2\right)= -\rho_1 k_1 c_1^2 \frac{j_0(X)}{j_1(X_0)} R\left(1-\frac{R_0}{R}+\frac{\dot{R}}{c_0}\right)
-4(\mu_0-\mu_1)\frac{\dot{R}}{R}-2\sigma\left(\frac{1}{R}-\frac{1}{R_0}\right)-p_A(t). 
\end{equation}

\section{SI.2 Effect of a neighboring rigid wall}

The equations of SI.1 can be extended to include the acoustic interaction with a rigid wall through the addition of a pressure term $P_r$ representing the reflected scattering or a droplet position at the mirrored location:
\begin{equation}
P_{r}=\rho_0 \frac{\partial}{\partial t} \frac{R^2\dot{R}}{2d},
\end{equation}
with $d$ the distance to the wall. Here we take $d=R$ for a droplet at the wall.
Then with $\rho_0(\frac{1}{2}\dot{R}^2+\frac{1}{2}R\ddot{R})$ this changes the left-hand side of Eq.~\ref{eq:pressure_diff_eq4} as follows:
\begin{equation}
\label{eq:pressure_diff_eq3}
\rho_0 \left(\frac{3}{2}R\ddot{R}+2\dot{R}^2\right)= -\rho_1 k_1 c_1^2 \frac{j_0(X)}{j_1(X_0)} R \left(1-\frac{R_0}{R}+\frac{\dot{R}}{c_0}\right)
-4(\mu_0-\mu_1)\frac{\dot{R}}{R}-2\sigma\left(\frac{1}{R}-\frac{1}{R_0}\right)-p_A(t).   
\end{equation}

\newpage

{
	
	\section{SI.3 Data for the physical parameters of water and PFP}
	
	\begin{table}[htp]
		{
			\caption{Physical parameters for water and PFP at standard temperature and/or standard pressure~\cite{NIST} (* the polar molecular interactions are neglected).  }
			
			\begin{center}
				\begin{tabular}{p{0.2\textwidth}>{\centering}p{0.2\textwidth}>{\centering}p{0.2\textwidth}>{\raggedright}p{0.1\textwidth}>{\raggedright\arraybackslash}p{0.1\textwidth}}
					& & & &\\
					\hline
					\textbf{parameter}				& \textbf{water} & \textbf{PFP}	& \textbf{unit} & \textbf{reference}	\\
					\hline\hline
					surface tension			& 72	& 10	& mN/m 	&~\cite{NIST,Kandadai2010}\\
					\hline
					interfacial tension*	& \multicolumn{2}{c}{62}	& mN/m 	&~\cite{NIST,Kandadai2010}\\
					\hline
					density					& 1000	& 1630	& kg/m$^3$ &~\cite{NIST}\\
					\hline
					viscosity				& 1.0	& 0.5	& mPa$\cdot$s &~\cite{NIST,Morgado2011}\\
					\hline
					speed of sound			& 1490	& 458	& m/s 	&~\cite{NIST}\\
					\hline
					boiling point			& 373	& 303	& K 	&~\cite{NIST}\\
					\hline
					vapor pressure			& 2.3	& 77	& kPa 	&~\cite{NIST}\\
					\hline
				\end{tabular}
			\end{center}
			\label{default}
		}
	\end{table}
	
}
\newpage

\section{SI.4 On the comparison with superharmonic focusing}

\begin{figure*}[b]
	\includegraphics[width=0.8\textwidth]{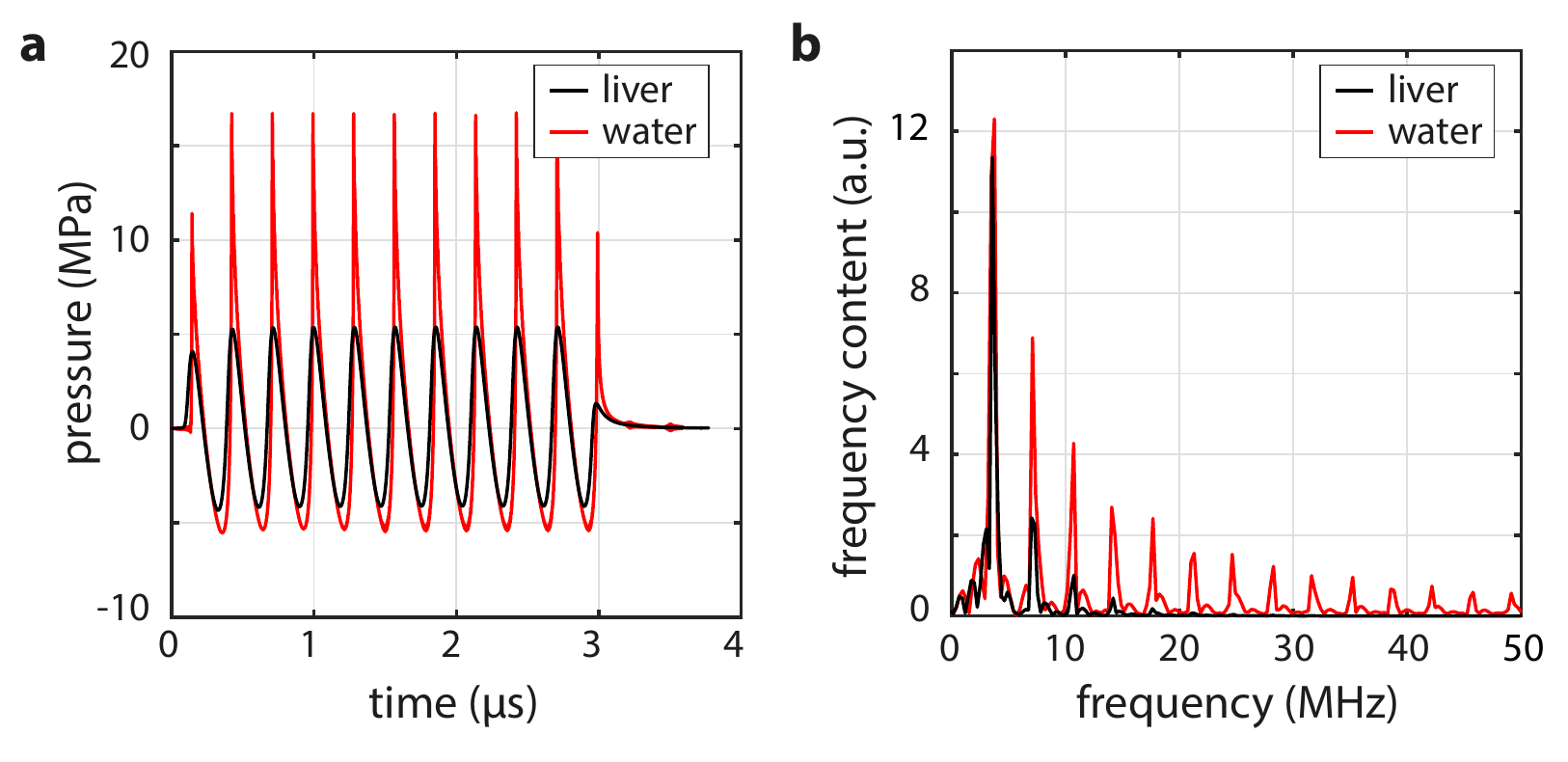}
	\caption{(a) Acoustic pressures resulting from non-linear propagation in water and liver tissue for identical pressure conditions and (b) corresponding Fourier spectra. The frequency of the initial acoustic wave is 3.5~MHz}
	\label{SI:2}
\end{figure*}

\begin{figure*}[t]
	\includegraphics[width=0.8\textwidth]{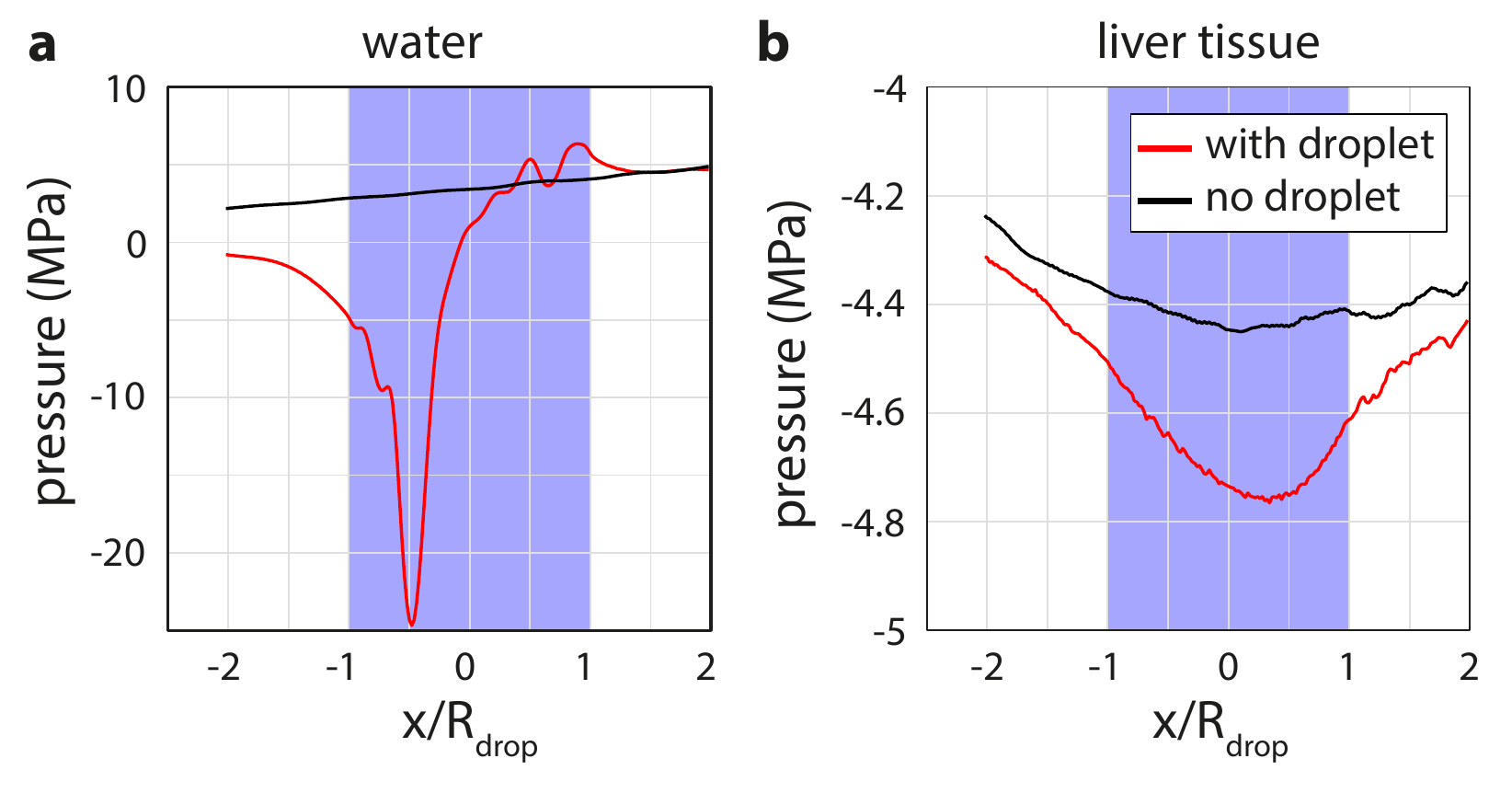}
	\caption{Minimum pressure distribution in a 10 $\mu$m radius droplet insonified with the waveforms displayed in Fig.~\ref{SI:2} in water (a) and in liver tissue (b).}
	\label{SI:3}
\end{figure*}

Nonlinear propagation highly depends on the properties of the medium and effectively results from a competition between nonlinearity of the medium and attenuation. For this reason, nonlinear propagation is much stronger in water that in soft human tissue, see Fig.~\ref{SI:2}. Since nonlinear propagation is the source of the superharmonic focusing effect (demonstrated in Fig.~\ref{SI:3}), it is unlikely to find a practical application in vivo. 

\begin{figure*}
	\includegraphics[width=0.7\textwidth]{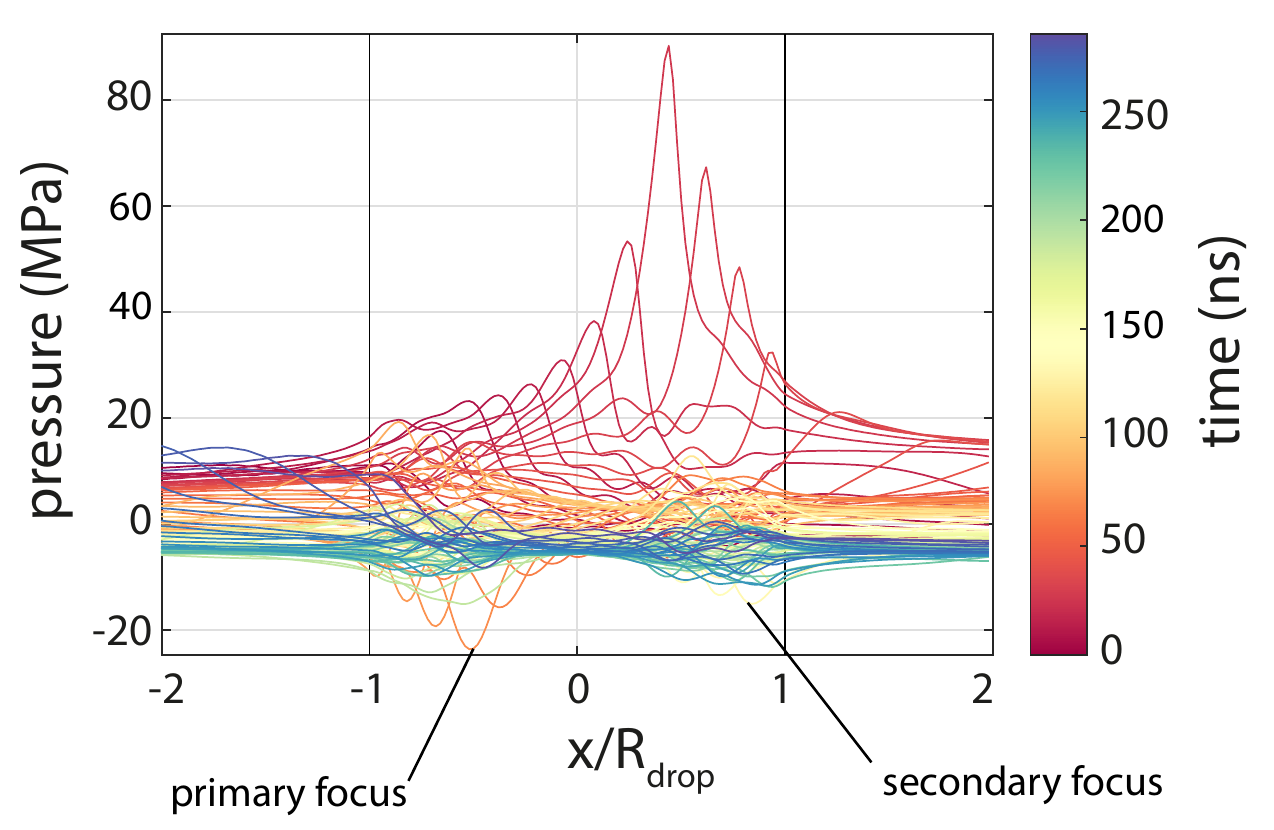}
	\caption{Pressure distribution in a 10$\mu$m radius droplet insonified with the waveform of Fig.~\ref{SI:2} in water, showing the primary proximal focus, resulting from a negative reflection on the distal side of the droplet, and the secondary focus, distal and less intense.}
	\label{SI:4}
\end{figure*}

Superharmonic focusing results from the coupling of the higher harmonic frequencies (generated during non-linear propagation) within the droplet, resulting, first, in a distal focus of positive pressure. The reflection of the wave at the distal droplet wall (with a negative reflection coefficient in pressure) leads to a focus of negative pressure on the proximal side of the droplet, as observed in~\cite{Shpak2014}. Successive reflections within the droplet generate a secondary negative focus, less intense at the distal side of the droplet (see Fig.~\ref{SI:4}). In this sense, the superharmonic focusing mechanism for ADV physically differs from the proposed resonance mechanism: since the typical timescale of the short waves interacting with the droplet in superharmonic focusing is at least an order of magnitude shorter that the frequency of the carrying wave, superharmonic focusing is effectively a repeated transient effect and does not lead to the accumulation of energy that occurs during resonant driving of the droplet.

\begin{figure*}
	\includegraphics[width=1\textwidth]{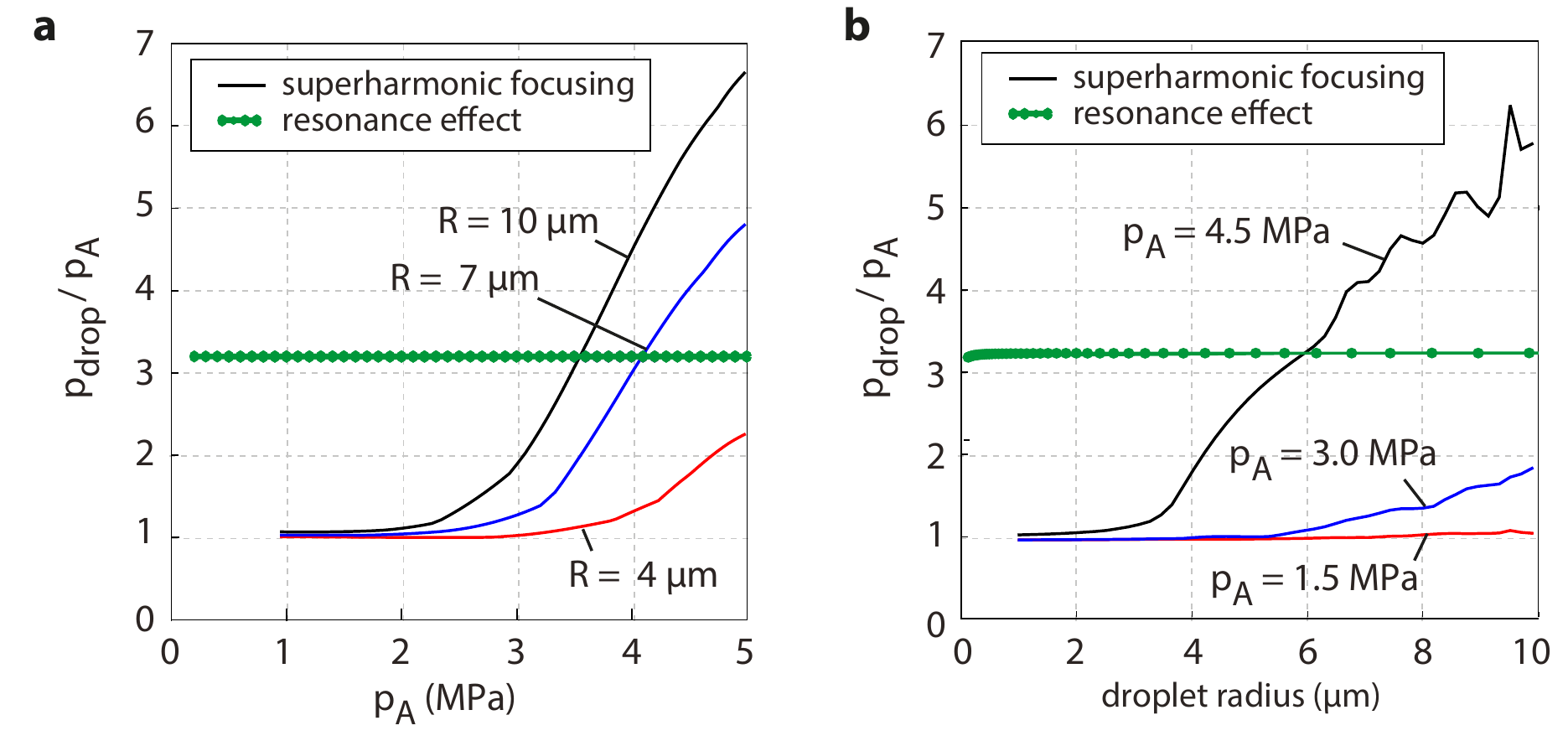}
	\caption{(a) Comparison of the efficiency of the superharmonic focusing effect and the resonance mechanism for an increasing incident pressure. (b) Comparison of the efficiency of the superharmonic focusing effect and resonance mechanism for varying droplet radius. The superharmonic focusing is calculated in water, for a fundamental frequency of 3.5 MHz.}
	\label{SI:5}
\end{figure*}

The proposed resonance mechanism requires high frequencies ($kR \sim 1$) but keeps a constant efficiency thoughout the parameter space, unlike superharmonic focusing, see Fig.~\ref{SI:5}. The logic, however, aligns well with the recent technological developments in the field of intravascular ultrasound (IVUS) that make use catheters to locally administer high frequencies ($>$ 50~MHz).

\clearpage
\newpage
\section{SI.5 Numerical simulations in k-wave}

\begin{figure*}[b]
	\includegraphics[width=0.7\textwidth]{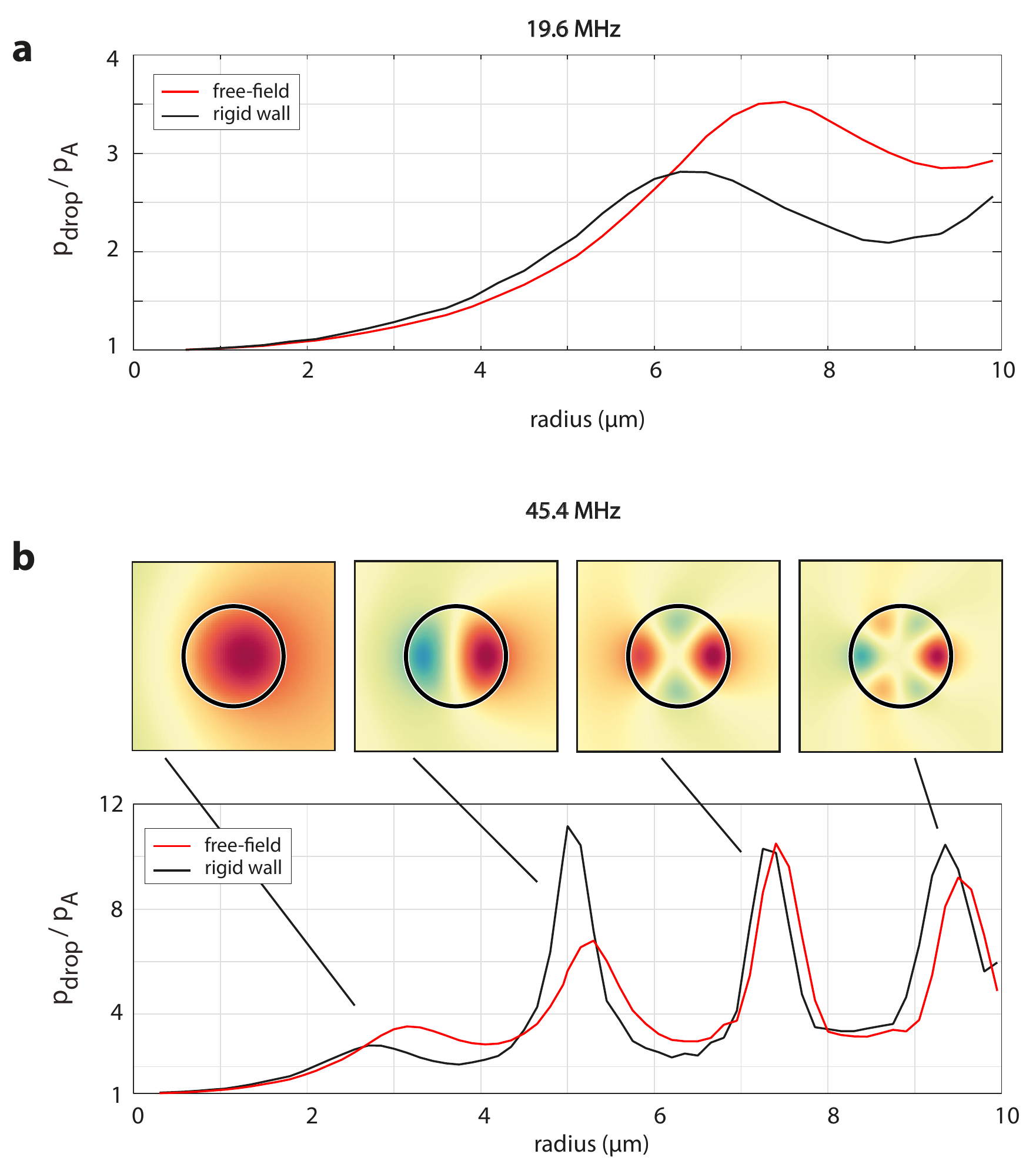}
	\caption{(a) Pressure amplification factor in the droplet for a driving frequency of 19.6~MHz. (b) Pressure amplification factor in the droplet for a driving frequency of 45.4~MHz with on top the simulated pressure fields in and around the droplet showing the fundamental and higher-order resonance modes.} 
	\label{SI:1}
\end{figure*}

Acoustic simulations were performed in k-wave (an open source Matlab toolbox)~\cite{Treeby2010a} in an axisymmetric configuration. Simulations, executed on GPUs, were run for varying droplet size, with and without the presence of a rigid substrate, and for both frequencies: 19.6~MHz and 45.4~MHz. This resulted in a set of modes and amplification factors, each shown in Fig.~\ref{SI:1}. 
\clearpage
\newpage

\section{SI.6 Derivation of the average nucleation rate}

Classical nucleation theory \cite{Caupin2006, Karthika2016} dictates that the nucleation rate $\Gamma$ is governed by a set of physical parameters as follows:
\begin{equation} \label{eq:nuc_rate}
\Gamma 	\propto exp{\left(  -\dfrac{16\pi}{3k_BT_{amb}} \dfrac{\sigma_{\textup{PFP}}^3}{\left( p_v - p_{atm} - p_{drop} \right) ^2}   \right)}.
\end{equation}
Here, $\sigma_{\textup{PFP}}$ is the surface tension of the liquid, $T_{amb}$ the ambient temperature and $p_{atm}$ the ambient pressure, $p_v$ is the vapor pressure of the liquid and $k_B$ is the Boltzmann constant. $p_{drop}$ is the pressure in the droplet and is given by Eq.~(\ref{eq:pressure_time}):
\begin{equation} \label{eq:pressure_time_2}
p(r,t)=p(r,0) - \rho_1 k_1 c_1^2 \frac{j_0(x) }{j_1(X_0)}(R-R_0), 
\end{equation}
with $x=\omega r/c_1$. Considering the large peak negative pressures typically required to induce cavitation, $\left( p_v - p_{atm}\right)/p_{drop} \ll 1$,   Eq.~(\ref{eq:nuc_rate}) can be reduced to:
\begin{equation} \label{eq:nuc_rate_simp}
\Gamma 	\propto exp{\left[  -\dfrac{C_0}{(p(r,t)^2}   \right]},
\end{equation}
with $C_0 = \dfrac{16\pi\sigma_{\textup{PFP}}^3}{3k_BT_{amb}}$. \\

$p(r,t)$ can be rewritten as the product of a radial pressure term $p(r)$, containing the $j_0(x)$ spherical Bessel function, with peak negative pressure $P_{min}$ at $r=0$, and a time-dependent pressure term $p(t)$ that oscillates with $cos(\omega t)$. Equation~(\ref{eq:nuc_rate_simp}) can now be volume and time integrated (during the negative pressure phase of the ultrasound cycle) to determine the average nucleation rate:
\begin{equation} \label{eq:av_nuc_rate_int1}
	<\Gamma> ~\propto \int^{R_{0}}_{0} \int^{T/4}_{-T/4} 4\pi r^2 exp{\left[  -\dfrac{C_0}{(p(r)^2cos^2(\omega t)}   \right]} dt~dr,
\end{equation}
with T the period of the ultrasound driving.
This integral is further simplified by symmetry of the integral around peak negative pressure (PNP) at $t=0$ (where $cos^2(\omega t)$ = 1) and substituting $y=cos(\omega t)$. Using $dt=\dfrac{1}{\omega}\dfrac{1}{\sqrt{1-y^2}}dy$ Eq.~(\ref{eq:av_nuc_rate_int1}) then becomes:
\begin{equation} \label{eq:av_nuc_rate_int2}
	<\Gamma> ~\propto \int^{R_{0}}_{0} r^2 \int^{1}_{0}  \frac{exp({-C_1/y^2})}{\sqrt{1-y^2}}dy~dr,
\end{equation}
with $C_1 = \dfrac{C_0}{p(r)^2}$. \\

\begin{figure*}[b]
	\includegraphics[width=1\textwidth]{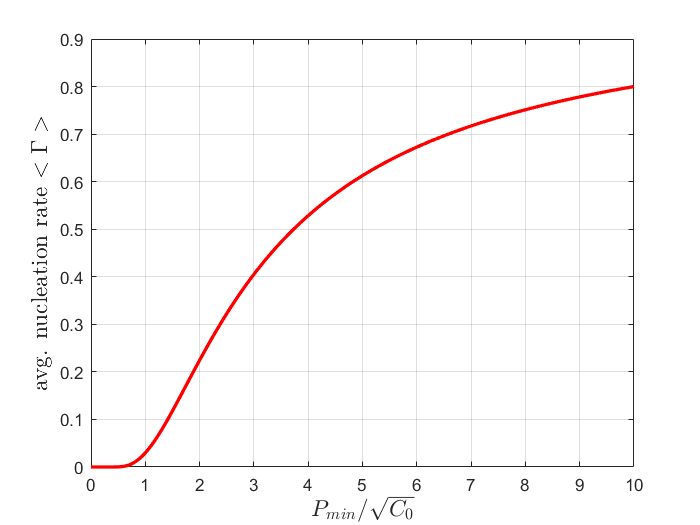}
	\caption{The numerical solution for the average nucleation rate $<\Gamma>$ as a function of the PNP pressure.}
	\label{SI:7}
\end{figure*}

The solution of the time-integral is given by an error function ($erf$):
\begin{equation}
\int^{1}_{0}  \frac{exp({-C_1/y^2})}{\sqrt{1-y^2}}dy = 1-erf{\left(\sqrt{C_1} \right)},
\end{equation} 
and Eq.~(\ref{eq:av_nuc_rate_int2}) now writes:
\begin{equation} \label{eq:av_nuc_rate_int3}
	<\Gamma> ~\propto \int^{R_{0}}_{0} r^2 \left[1-erf{\left(\frac{\sqrt{C_0}}{p(r)} \right)}\right]dr.
\end{equation}
To our knowledge Eq.~(\ref{eq:av_nuc_rate_int3}) has no simple analytical solution and it can be solved numerically to give the result displayed in Fig.~\ref{SI:7}.

\newpage
\bibliography{../Bibliography_5}


\maketitle
\section*{Supplementary Movie 1}
Axisymmetric acoustic simulation in k-wave of a 45.4 MHz acoustic wave propagating (left to right) from a lithium niobate substrate  towards a 2.7 $\mu$m radius perfluoropentane droplet immersed in water. The droplet resonantly builds up pressure over the first few ultrasound cycles until it reaches a pressure amplitude that is $\sim$~2.8 times larger than that of the acoustic wave in water.

\begin{wraptable}{l}{0.5cm}
\begin{tabular}{l|l} 
\hline \hline
file name   & ADV\textunderscore resonance.mp4 \\
file size   & 13.6 MB \\
format.     & H.264 \\
resolution  & 626 $\times$ 526 \\
encoded fps & 30 \\
\hline
\end{tabular}
\label{tab:table1}
\end{wraptable}